\documentclass[aps,prb,twocolumn,superscriptaddress,showpacs,preprintnumbers,amsmath,amssymb,floatfix]{revtex4}

\usepackage{graphicx}
\usepackage{float}
\usepackage{bm}
\usepackage{epstopdf}
\usepackage{multirow}
\usepackage{hyperref}
\usepackage{color}
\hypersetup{ 
	colorlinks,
	linkcolor=blue,
	filecolor=black,
	urlcolor=black,
	citecolor=blue}

\begin{document}

\title{Damping of Landau levels in neutral graphene at low magnetic fields: A phonon Raman scattering study}

\author{F. M. Ardito}
\affiliation{``Gleb Wataghin'' Institute of Physics, University of Campinas - UNICAMP, Campinas, S\~ao Paulo 13083-859, Brazil}

\author{T. G. Mendes-de-S\'{a}}
\affiliation{Departamento de F\'{i}sica, Universidade Federal de Minas Gerais, Belo Horizonte, Minas Gerais 30123-970, Brazil}

\author{A. R. Cadore}
\affiliation{Departamento de F\'{i}sica, Universidade Federal de Minas Gerais, Belo Horizonte, Minas Gerais 30123-970, Brazil}

\author{P. F. Gomes}
\affiliation{``Gleb Wataghin'' Institute of Physics, University of Campinas - UNICAMP, Campinas, S\~ao Paulo 13083-859, Brazil}
\affiliation{Instituto de Ciências Exatas e Tecnol\'{o}gicas, Universidade Federal de Goi\'{a}s, Jata\'{i} 75801-615, Brazil}

\author{D. L. Mafra}
\affiliation{Departamento de F\'{i}sica, Universidade Federal de Minas Gerais, Belo Horizonte, Minas Gerais 30123-970, Brazil}

\author{I. D. Barcelos}
\affiliation{Departamento de F\'{i}sica, Universidade Federal de Minas Gerais, Belo Horizonte, Minas Gerais 30123-970, Brazil}

\author{R. G. Lacerda}
\affiliation{Departamento de F\'{i}sica, Universidade Federal de Minas Gerais, Belo Horizonte, Minas Gerais 30123-970, Brazil}

\author{F. Iikawa}
\affiliation{``Gleb Wataghin'' Institute of Physics, University of Campinas - UNICAMP, Campinas, S\~ao Paulo 13083-859, Brazil}

\author{E. Granado}
\affiliation{``Gleb Wataghin'' Institute of Physics, University of Campinas - UNICAMP, Campinas, S\~ao Paulo 13083-859, Brazil}

\begin{abstract}

Landau level broadening mechanisms in electrically neutral and quasineutral graphene were investigated through micro-magneto-Raman experiments in three different samples, namely, a natural single-layer graphene flake and a back-gated single-layer device, both deposited over Si/SiO$_2$ substrates, and a multilayer epitaxial graphene employed as a reference sample. Interband Landau level transition widths were estimated through a quantitative analysis of the magnetophonon resonances associated with optically active Landau level transitions crossing the energy of the $E_{2g}$ Raman-active phonon. Contrary to multilayer graphene, the single-layer graphene samples show a strong damping of the low-field resonances, consistent with an additional broadening contribution of the Landau level energies arising from a random strain field. This extra contribution is properly quantified in terms of a pseudomagnetic field distribution $\Delta B=1.0-1.7$ T in our single-layer samples.
\end{abstract}


\maketitle

\section{Introduction}

The singular half-integer quantum Hall effect in graphene is a direct consequence of the characteristic Landau levels (LLs) predicted by the Dirac equation. Although sharp levels are required to reinforce the manifestation of this effect, limited information on the most relevant mechanisms leading to broadening of quantized electronic levels in graphene samples is presently available. The quantized energies for the linear electronic bands around the Dirac points in graphene are $E_{n}= sgn (n)v_{F}\sqrt{2e\hbar B |n|}$,\cite{Castro Neto} where the index $n=0$, $\pm 1$, $\pm2$, ..., $v_F$ is the Fermi velocity, and $B$ is the magnetic field perpendicular to the carbon sheet. Ideal graphene, i.e., a perfectly flat, isolated, defect-free and strain-free layer, is expected to show sharp LLs at low temperatures, with small intrinsic broadening ($\delta E_n \lesssim 1$ meV for $B=4$ T) due to carrier-carrier, carrier-light, and carrier-phonon interactions. \cite{Funk} On the other hand, real samples show imperfections that are characteristic of the sample production method, leading to LL broadening and consequent damping of the effects associated with the Dirac equation. A proper understanding of the main mechanisms of extrinsic LL broadening is therefore desirable and should help in the quest for optimized graphene samples with reinforced quantum relativistic effects.

As detailed below, the mechanisms of LL broadening may be pinned down by a quantitative analysis of the {\it B} dependence of the LL width. Such information can be achieved  by direct observations of the LLs by scanning tunneling spectroscopy, \cite{Li_STS,Mayer_STS} infrared absorption,\cite{SadowskiIR,JiangIR,OrlitaIR,OrlitaIR2} and Raman scattering. \cite{PhysRevLett.107.036807,FaugerasEPL,Faugeras2015} Alternatively, the broadening of LLs may be conveniently studied by an analysis of phonon Raman scattering, which is a versatile and widespread technique that probes structural and electronic properties of graphitic samples.\cite{Malard,Pimenta} In fact, electron-phonon interaction in graphene leads to magnetophonon resonances (MPRs) when the energy of an optically active LL transition obeying $|n|-|m|=1$ crosses the energy of the $E_{2g}$ Raman-active phonon,\cite{ando2005electronic,tando,Goerbig} causing oscillations of the phonon energy and linewidth. Several works have reported the MPRs from Dirac fermions in graphene and graphitic samples.\cite{PhysRevLett.107.036807,faugerasfit,PhysRevLett.105.227401,PhysRevB.85.121403,Goler20121289,kim2013filling,leszczynski2014electrical,neumann2015magneto} For electrically neutral graphene, the MPRs are described by  \cite{faugerasfit}
\begin{equation}
\widetilde{\epsilon}^2 - \epsilon_{0}^2= 4 \lambda \epsilon_{0} e\hbar B  v_{F}^{2} \sum\limits_{k=0}^{\infty}\left[ \frac{ T_{k}}{\left( \widetilde{\epsilon}+\imath \delta_k \right) ^2 - T_k^2} + \frac{1}{T_k} \right ]
\label{eq1}
\end{equation}
where $\epsilon_{0}$ stands for the phonon energy in the absence of magnetic field; $\lambda$ is the electron-phonon coupling parameter; and $T_{k}=v_{f}\sqrt{2e\hbar B} \left( \sqrt{k} + \sqrt{k+1} \right)$ describes the energy of interband LL transitions, with index $k$, for $\left| n \right| - \left| m \right| = 1$, in which $n$ and $m$ are labels for the initial and final Landau levels involved in the transition; and $\delta_k$ represents the LL broadening parameter, which is of particular significance to the present work. The real and complex parts of $\widetilde{\epsilon}=\epsilon-\imath \Gamma$ yield the phonon energy and broadening from the electron-phonon coupling, respectively. Comparison between Eq. (\ref{eq1}) and the $E_{2g}$ phonon energy and linewidth experimentally obtained as a function of magnetic field allows one to extract sample-related parameters such as $v_F$, $\lambda$, and $\delta_k$. In this work, the MPRs of three distinct graphene samples, namely, a multilayer epitaxial graphene (MEG), a single-layer graphene (S1) deposited over a SiO$_2$ substrate, and a back-gated single-layer device (S2), were quantitatively analyzed by means of Eq. \ref{eq1} in order to extract information on the broadening $\delta_k$ parameter. The distinct behavior of $\delta_k$ as a function of the resonance index $k$ found for MEG, S1, and S2 samples allowed us to identify an additional LL broadening mechanism for single layers that is not present in MEG, associated with random strain field.

\section{Experimental Details}

The MEG sample was obtained by decomposition of the carbon face of $4H$-SiC$(000\overline{1})$ substrate in argon atmosphere. Sublimation time was 60 min at $T = 2048$ K. Further details on the preparation and characterization of this sample by Raman scattering, atomic force microscopy, and grazing-incidence x-ray diffraction are given in a previous work.\cite{mendes2012} MEG samples are known to present weak electronic coupling between graphene layers, \cite{FaugerasAPL} also showing very small carrier concentrations of the order of $10^{10}$ cm$^{-2}$ or lower.\cite{SadowskiIR,OrlitaIR} The single layer S1 flake was produced using conventional mechanical exfoliation of natural graphene deposited over the $300$-nm SiO$_2$ layer of a Si substrate. The single-layer S2 device was prepared using the standard scotch-tape method deposited over the $285$-nm SiO$_2$ grown on top of highly $p$-type doped Si wafers. Metallic contacts Cr/Au ($1/40$~nm) were patterned by standard electron-beam lithography and thermal metal deposition. To remove polymer residues remaining from the lithography processes and avoid external doping,\cite{cadore2016} the device was submitted to a final thermal annealing step at $T=350^{\circ}$C for $3$~h under H$_2$/Ar ($300/700$~sccm). The device has a two-terminal geometry [see Fig.~\ref{fig:SLG2_char}(a) below], and electronic measurements were performed using a standard lock-in technique, applying a current bias of $I_{SD}=100$~nA at $17$~Hz through the graphene channel. For the measurements as a function of the back-gate voltage $V_{BG}$, the doped Si substrate was used as the back-gate electrode, and we worked with safe limits of $V_{BG}=±60$~V, from which we measured our devices during days without any leakage current through the dielectric. 
The carrier charge mobility was determined according to the expression $\mu=(\frac{L}{WC_G})\frac{dG}{dV_{BG}}$, in which $L$ and $W$ are, respectively, the length and width of the graphene channel, $C_G$ is the capacitance per unit of area, and $G$ is conductance. 

The micro-Raman and electrical experiments under magnetic fields were performed using a 15-T optical magnetocryostat. The sample, which was fixed to $xyz$ piezoelectric stages, and objective lens were immersed in a He gas or superfluid environment. The magnetic field was applied perpendicularly to the sample surface. The elastic component of the scattered light was rejected by an edge filter. Some details of the setup are specific for the experiment on each sample. The experiments on the S1 and MEG samples were performed using a $488$-nm Ar-ion laser, while a $532$-nm solid-state laser was employed for the experiment on S2. For sample S1, a single $1200$~g/mm grating spectrometer with a Peltier-cooled CCD detector was employed; we used a $40\times$ objective lens with a $200$-$\mu$m working distance, resulting in an $\sim 3.5$~$\mu$m focal spot diameter. For the experiment on the MEG and S2 samples, a single $1800$~g/mm grating spectrometer coupled with a liquid-nitrogen-cooled charge coupled device detector was employed; the laser was focused using a $50 \times$ objective lens, with a 7-mm working distance and a spot size of $\sim 2.5$ $\mu$m. A $200$-$\mu$m-diameter optical fiber, which works as a confocal configuration, was used to transport the Raman signal to the entrance of the spectrometer for the experiment on the MEG sample. 

\section{Results and analysis}

\subsection{Multilayer epitaxial graphene}

\begin{figure}
\includegraphics[width=0.5\textwidth]{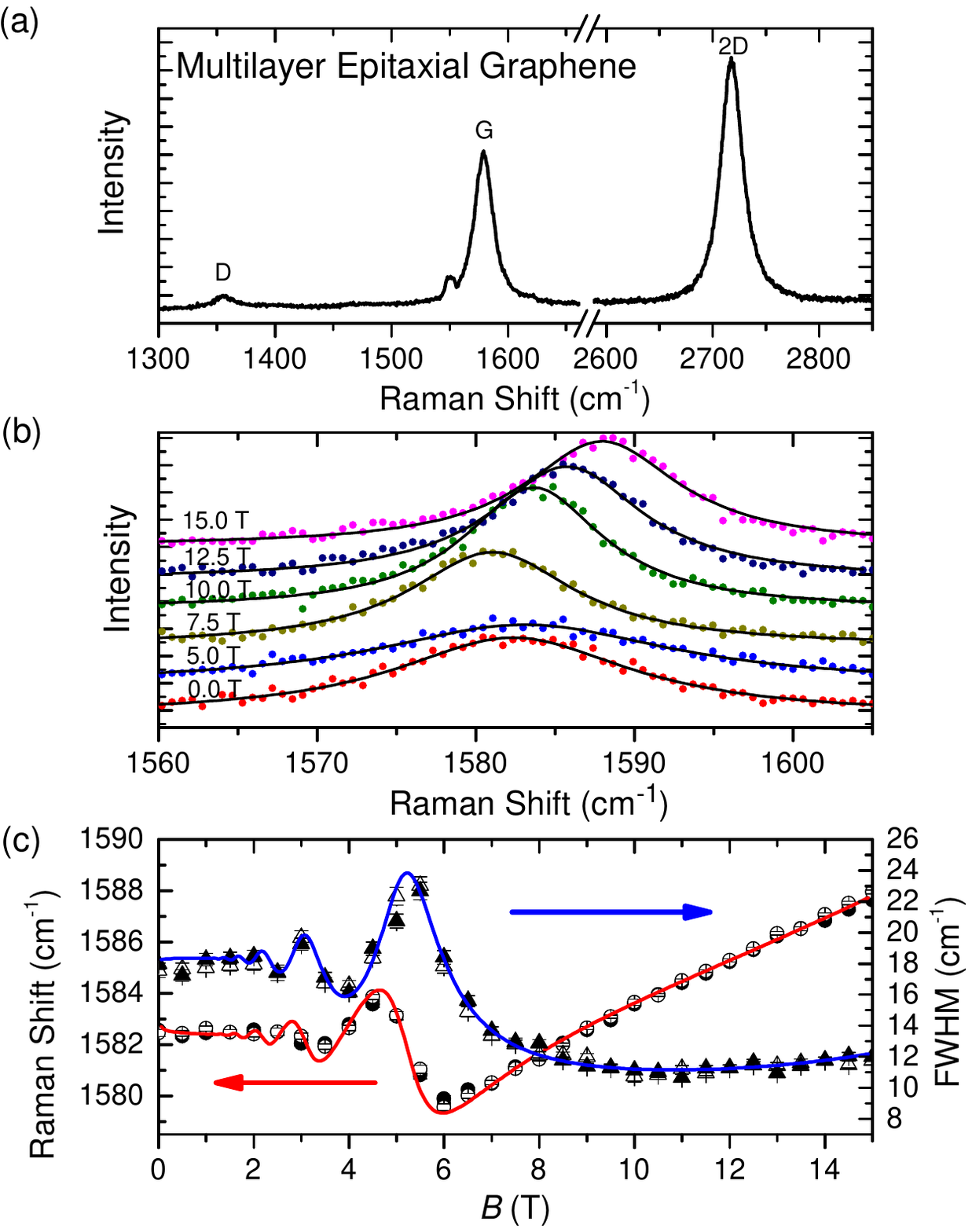}
\caption{\label{MEG} (a) Raman spectra of multilayer epitaxial graphene on SiC at 5.5 K and zero field. The characteristic graphene $D$, $G$, and $2D$ bands are indicated. The peak at $\sim 1555$~cm$^{-1}$ is a spurious signal due to parasitic scattering in the optical fiber. (b) $G$ band at selected magnetic fields and $T=5.5$ K; points represent experimental data, and solid lines are Lorentzian curve fittings. (c) $E_{2g}$ phonon energy and full width at half maximum (FWHM) as a function of magnetic field. Open and solid symbols represent two sets of data, taken on different spots of the sample, both showing a graphenelike single-peaked $2D$ band; solid lines represents a simulation according to Eq. (\ref{eq1}) using a single Landau level width $\delta = 17.6$ meV for all inter-LL transitions.}
\end{figure}

MEG samples are well known to show large magnetophonon resonance effects at relatively low fields,\cite{faugerasfit} therefore being appropriate to test the methodology employed here. Sample regions showing graphenelike Raman spectra with the sharpest $2D$ bands were chosen for our study. Figure \ref{MEG}(a) shows the Raman spectrum at 5 K with the characteristic $G$ and $2D$ bands, as well as the defect $D$ band, indicating a small but detectable degree of structural defects in this sample. A spectral interval near the $E_{2g}$ mode ($G$ band) is displayed in Fig.~\ref{MEG}(b) for selected applied magnetic fields, revealing a clear $B$ sensitivity. Single-Lorentzian fits were performed [solid lines in Fig.~\ref{MEG}(b)], and the peak energy and linewidth (FWHM) were extracted for two sets of data obtained at distinct spot positions on the sample, yielding reproducible oscillations with field that are signatures of the magnetophonon resonance in graphene [see Fig.~\ref{MEG}(c)]. \cite{faugerasfit} An excellent match with experimental data is obtained if Eq.(\ref{eq1}) is employed with the parameters $\epsilon_0=1581.7\:$~cm$^{-1}$, $v_{f}=0.985\times10^{6}\:$m/s, $\lambda=4.1\times10^{-3}$, and $\delta_k =17.6$ meV for all $k$ [solid lines in Fig. \ref{MEG}(c)]. A constant phonon linewidth contribution $\Gamma_{0}=8.2\:$~cm$^{-1}$, attributed to phonon decay processes not related to the electron-phonon coupling, was convoluted with $\Gamma$ to model the total $B$-dependent linewidth of the $G$ band. Overall, the extracted parameters are comparable to those previously reported for another MEG sample.\cite{faugerasfit}

\subsection{Single-layer graphene on SiO$_2$}

\begin{figure}
	\centering
		\includegraphics[width=0.5\textwidth]{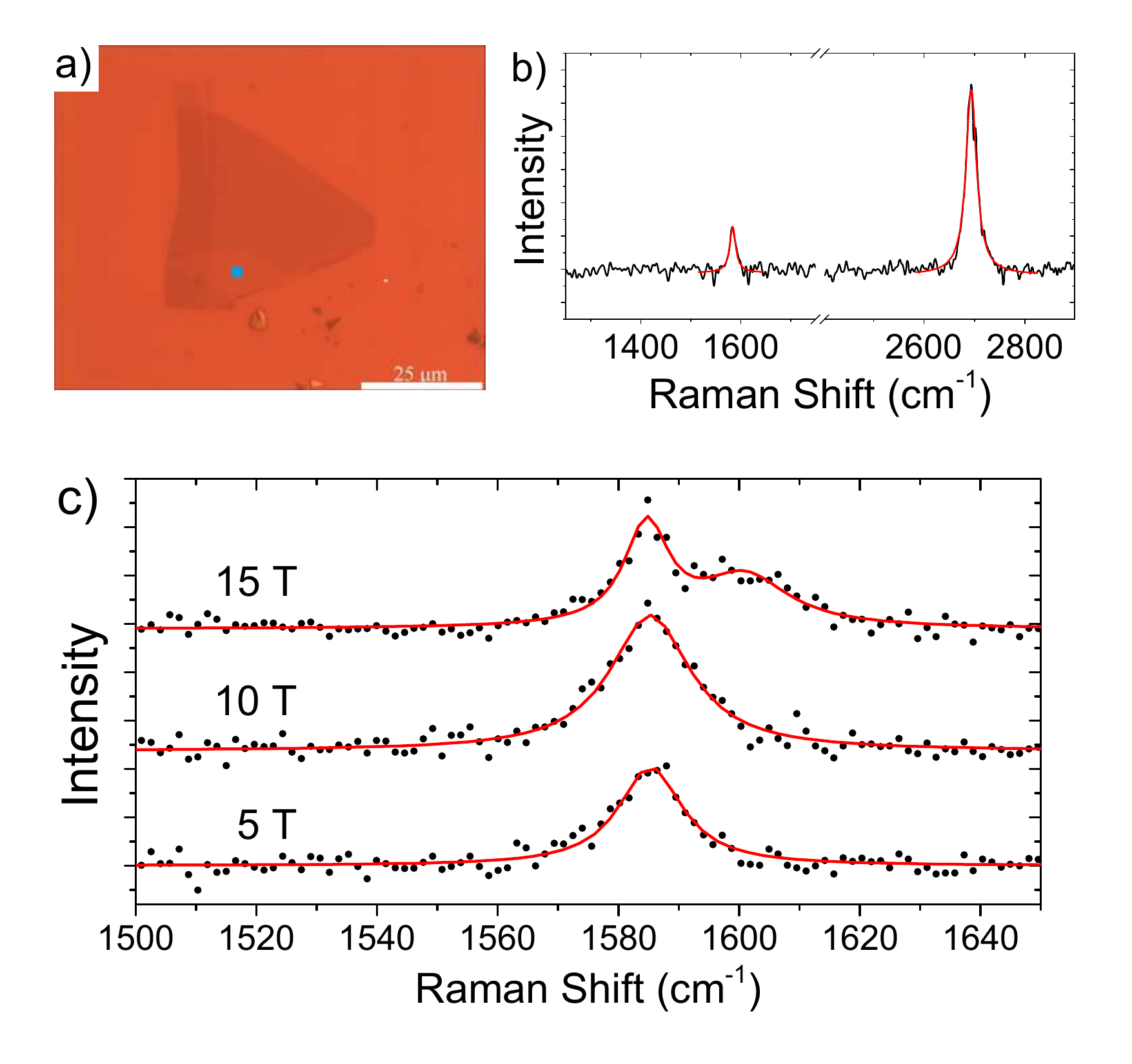}
\caption{\label{SLG} (a) Optical image of our exfoliated graphene sample, obtained using an orange filter. The blue circle indicates the approximate dimension and position of the laser spot focused into the single-layer graphene region S1. (b) Raman spectrum of S1 at room temperature. The $G$ and $2D$ bands are observed at 1584 and 2694 cm$^{-1}$ with full widths at half maxima of 14.3(9) and 25.4(3) cm$^{-1}$, respectively. The $2D/G$ peak area and peak height ratios are 7.2(5) and 4.1(3), respectively. (c) $G$ band of the single-layer sample at $T=5$ K and various magnetic fields. Solid lines in (b) and (c) are Lorentzian fits to the observed peaks.}
\end{figure}

\begin{figure}
	\includegraphics[width=0.5\textwidth]{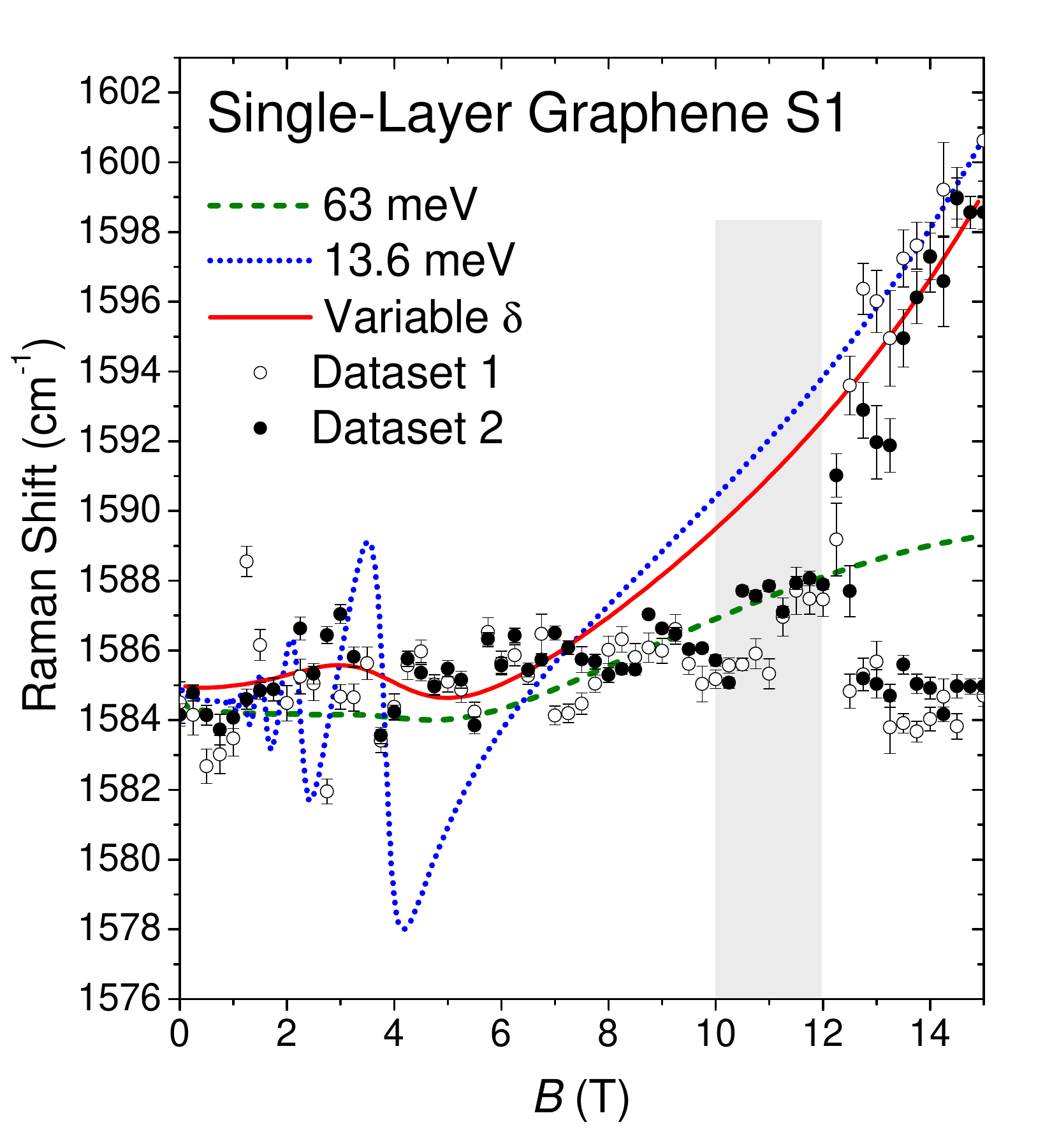}	
\caption{\label{SLGB} $G$-band central positions for the single-layer graphene sample S1, obtained with fits using a single Lorentzian below 12 T and two Lorentzians above 12 T [see also Fig. \ref{SLG}(c)] for data at $T=5$ K. The gray area marks an intermediate-field region where a single peak was employed in the fit, although a double-peak structure, not resolved in our data, is likely to be present. Empty and filled circles refer to data sets taken on two independent runs. Dotted blue and dashed green lines are the simulated magnetophonon resonance effect according to Eq. (\ref*{eq1}) using fixed Landau level broadening parameters $\delta=13.6$ and 63 meV, respectively. The solid red line shows the results of a simulation using Eq. (\ref*{eq1}) with the $\delta$ parameter dependent on $B$  using $\Delta B=1.7$~T and $\delta_0 = 6.3$ meV (see text).}
\end{figure}

Figure \ref{SLG}(a) shows the optical image of the S1 single-layer graphene flake. Figure \ref{SLG}(b) shows the Raman spectrum at room temperature with the characteristic $G$ and $2D$ bands and no sign of the defect-activated $D$ band. This result indicates the absence of structural defects within our sensitivity. The $G$-band Raman spectrum for selected magnetic fields and $T=5$ K is displayed in Fig. \ref{SLG}(c). This band clearly splits in two peaks above $\sim 12$ T, in line with previous reports. \cite{kim2013filling,Remi} Figure \ref{SLGB} shows the energy of the $G$ band at 5 K as a function of magnetic field. For $B \lesssim 12$ T, where a single $G$ band was observed within our resolution, no magnetophonon resonance could be detected, in stark contrast to the MEG sample. For $B>12$ T, one of the components of the split $G-$band remains at a nearly constant energy position, while a second component follows a preresonant behavior associated with the $k=1$ ($n=0 \leftrightarrow \pm 1$) inter-Landau-level transitions at $B_{res}^{k=1}=25-30$ T.\cite{kim2013filling} Note that the peak position of this field-dependent component follows a similar behavior found for the MEG sample in the higher-field regime above 12 T [see also Fig. \ref{MEG}(c)]. It is therefore evident from our results and from the literature \cite{kim2013filling,Remi} that single-layer graphene on SiO$_2$ is inhomogeneous and regions with two distinct behaviors with field are found within probed areas of a few square micrometers: (i) regions showing no observable magnetophonon resonance at all and (ii) regions showing clear manifestations of the main $k=1$ resonance.

\begin{figure}
	\centering
		\includegraphics[width=0.5\textwidth]{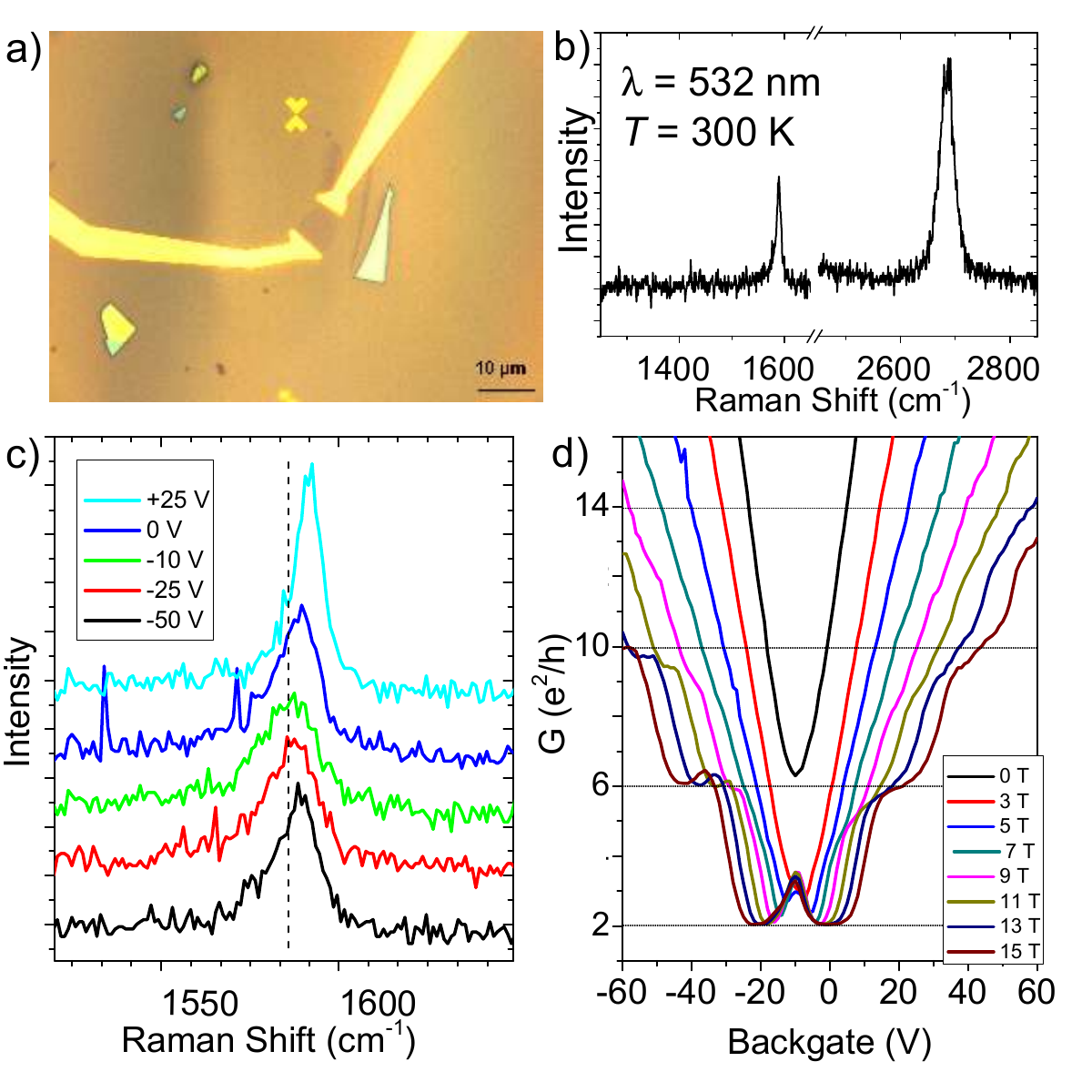}
	\caption{(a) Optical image of device S2. (b) Raman spectrum of S2 at ambient conditions. The $2D$ band is typical of single layers with one Lorentzian component and FWHM of $\sim 30$~cm$^{-1}$. (c) Gate response of the $G$ band for some back-gate voltages at room $T$. Minimum peak position occurs around $-10$~V. (d) Conductance as a function of gate voltage up to $15$~T at $T=10$ K. The minimum conductivity occurs at about $-10$~V for $B=0$ T, and plateaus are present at high $B$.}
	\label{fig:SLG2_char}
\end{figure}

The large contrast of the $G$-band behavior with field for the MEG and S1 samples is remarkable. Particularly, the absence of the low field ($\lesssim 8$ T) magnetophonon oscillations for the S1 sample does not seem to originate from structural defects since the defect $D$ band is present only in the MEG sample. To proceed, we must exclude the possibility of a small natural doping of the S1 sample causing a Pauli blocking of the observable LL transitions. A combined analysis of the peak intensities,\cite{Das} areas, \cite{Basko} positions, and linewidths \cite{Das,Pisana} of the $G$ and $2D$ bands extracted from the room-temperature Raman spectrum of the S1 sample reveals an electrically neutral graphene within experimental error [$n_0=(0 \pm 1) \times 10^{12}$ cm$^{-2}$]. However, since the uncertainty on $n_0$ is relatively large, we carried out a Raman investigation of the S2 sample, which is a back-gated device in which the Fermi level $E_F$ can be tuned [see Fig. \ref{fig:SLG2_char}(a)]. Figure ~\ref{fig:SLG2_char}(d) shows the electrical resistance of this sample as a function of $V_{BG}$ for magnetic field varying from $B=0$~T to $B=15$~T, showing the expected Landau levels at filling factors $\nu=±2,±6,±10…$ at high fields. The maximum mobility obtained for this device was $5000$~cm$^2$/Vs at $B=0$~T. Note that the neutrality point is reached by applying a back-gate voltage of $V_{BG} \sim -10$~V, indicating a small $n$-type natural doping. The transport asymmetry between electrons and holes that appear in two-probe measurements is believed to be responsible for small deviations from the expected conductance plateaus $G=νe^{2}/h$. Figure \ref{fig:SLG2_char}(b) shows the Raman spectrum of S2 at $B=0$ T, $T=300$ K, and null gate voltage, where the defect $D$ band is again absent, attesting to the good structural quality of this sample. The $G$ band at $B=0$ T, $T=300$ K, and various gate voltages is shown in Fig. \ref{fig:SLG2_char}(c). This band shows a clear dependence on $V_{BG}$, with maximum linewidth and minimal central energy at $V_{BG} \sim -10$ V, i.e., at the neutrality point ($E_F=0$) shown by transport measurements at the same experimental conditions. This is consistent with results shown in the literature.\cite{Pisana}

\begin{figure}
	\centering
		\includegraphics[width=0.5\textwidth]{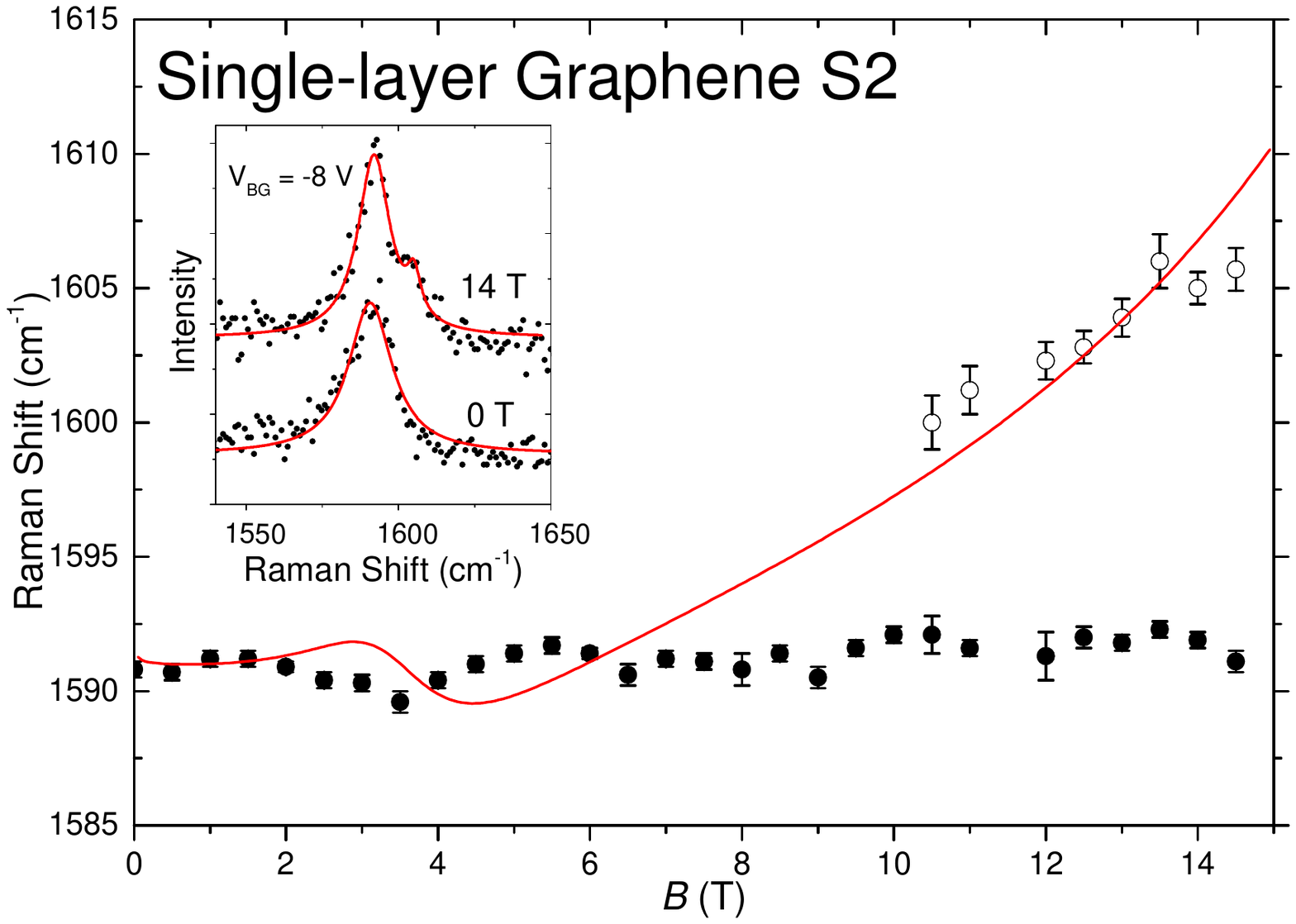}
	\caption{$G$-band energy as a function of magnetic field for sample S2 at $T=10$ K and $V_{BG}=-8$ V (neutrality point). The red line is a simulation including a pseudomagnetic field distribution $\Delta B = 1.0$ T. The inset shows the $G$ band at $B=0$ and $14$~T, where a two-peak line shape is evident.}
	\label{fig:SLG2_sim}
\end{figure}

The inset of Fig. \ref{fig:SLG2_sim} shows the $G$ band of sample S2 at $T=2$ K and $B=0$ and $14$ T, taken with $V_{BG} = -8$ V, which was the neutrality point for the conditions of this measurement at low $T$. As observed for sample S1 [see Fig. \ref{SLG}(b)], this band splits into two peaks at high $B$. Figure \ref{fig:SLG2_sim} shows the $B$-dependence of the energy central position of the observed peaks at the $G$ band of the S2 device. Remarkably, the results for the S2 $G$ band with $E_F=0$ are very similar to those acquired for the unprocessed sample S1 (see Fig. \ref{SLGB}), demonstrating that the absence of the low-$B$ MPR in single-layer graphene deposited on SiO$_2$ substrates is not explained by natural doping.

Insight into the damping of the low-field resonances for single layer graphene is gained by an analysis of Eq. (\ref{eq1}) as a function of the $\delta_k$ parameters related to the LL transition widths. In Fig.~\ref{SLGB}, the dashed and dotted lines show the calculated $B$ dependence of the $G$ band central wave number for two selected values of $\delta$, assumed so far to be the same for all transition indexes $k$. In these simulations, the parameters $\epsilon_0=1582\:$~cm$^{-1}$, $v_{F}=1.15\times10^{6}\:$m/s, $\lambda=5.5\times10^{-3}$ were employed. The different Fermi velocity $v_F$ for S1 with respect to the MEG sample is consistent with a previous observation of sample-dependent $v_F$ due to different strengths of electron-electron interactions\cite{Faugeras2015} for S1 and MEG samples. For $\delta=13.6$ meV, the high-$B$ preresonant behavior observed for one of the $G-$band components is captured. However, if the same $\delta$ is used for the other LL transitions, the resonance at 3.8 T associated with the $n=-1 \rightarrow 2$ and $n=-2 \rightarrow 1$ LL transitions remains prominent and would be clearly visible within our resolution. If, on the other hand, a much larger $\delta=63$ meV is employed in the simulations, all the magnetophonon resonances are washed out, including the observed preresonant behavior in the field range $B>$12 T. We conclude that, while the $B-$independent component of the $G$ band observed in single-layer graphene on SiO$_2$ may be attributed to sample regions showing a very large $\delta$ value, the behavior observed for the $B-$dependent $G$-band component cannot be explained by single-$\delta$ magnetophonon resonances. In fact, a much larger $\delta_k$ for the low-field resonances ($k \geq 2$) with respect to the main one ($k=$1) is necessary for Eq. (\ref*{eq1}) to capture the observed behavior of the $B-$dependent $G$-band component of our single-layer graphene samples.

\section{Discussion}

Our experimental data indicate that for MEG the LL broadening parameter $\delta_k$ in the magneto-phonon resonance [Eq. \ref*{eq1}] is actually independent of $k$, while for single-layer samples $\delta_k$ increases with $k$. We proceed with a discussion on the significance of these observations. As mentioned above, the manifestations of MPRs in the Raman spectra occur at fields $B_{res}^k$ where the interband LL energy difference equals the $G$-band energy, i.e., $E_{G-band}=v_{F}\sqrt{2e\hbar B_{res}^k}(\sqrt{|n|}+ \sqrt{|n+1|})$. In other words, in this experiment distinct LL transitions are probed at the same energy but different magnetic fields. The observation of a LL broadening parameter  $\delta$ that is independent of the transition index $k$ indicates that the LL width is proportional to its energy. Since the LL energy is in turn proportional to $\sqrt{B}$, this conclusion is consistent with direct measurements of LL widths as a function of $B$ for MEG samples, where a $\sqrt{B}$ dependence is found for the LL widths. \cite{OrlitaIR,OrlitaIR2} Recent theoretical work analyzed different microscopic mechanisms of LL broadening and attributed this $\sqrt{B}$ dependence to an extrinsic mechanism involving scattering of the charge carriers by impurities.\cite{Funk} In fact, intrinsic mechanisms such as scattering from carrier-carrier, carrier-light, and carrier-phonon interactions cannot explain the relatively large LL broadening observed for MEG ($\delta_k / E_k = 0.09$ for our sample). We should mention that an additional extrinsic mechanism of LL broadening involving fluctuations of $v_F$ from layer to layer may also lead to the same behavior with constant $\delta_k / E_k$ ratio since the LL energy is proportional to $v_F$. In fact, it is well established that $v_F$ is dependent on a residual interaction with the substrate or neighboring graphene layers, reaching maximum values for suspended single layer graphene samples,\cite{Hwang} making Fermi velocity fluctuations a plausible source of LL broadening in MEG.
 
For single layers, it is evident that an additional extrinsic mechanism must be present to account for the index-dependent broadening that washes out the resonances with $k \geq 2$. We suggest that such a mechanism is associated with strain fluctuations. In fact, while for MEG samples the graphene layers are self-protected, strain fluctuations associated with corrugation of single-layer graphene may be significant. In the absence of a complete microscopic theory that takes into account the effect of inhomogeneous strain in the LLs of graphene, we propose a phenomenological approach that seems to capture the essential physics. It has been shown that strain leads to a discretization of the electronic levels in graphene that is similar to the effect of an external magnetic field.\cite{Castro Neto,andoanomaly,PhysRevLett.97.196804} Since strain in single-layer graphene tends to be inhomogeneous, it is expected that a distribution of pseudo-magnetic fields takes place, which would introduce a certain standard deviation $\Delta B$ in the effective magnetic field and lead to an obvious pathway to LL broadening. Quantitatively, one would have $\Delta E_n/E_n = \Delta B / 2 B$.

In order to verify if the broadening mechanisms indicated above are consistent with our observations in S1 and S2, simulations of the $E_{2g}$ phonon energy and linewidth according to Eq. (\ref{eq1}) were performed considering the Lorentzian-convoluted parameter $\delta_k = \delta_0 + (\Delta B / 2 B_{res}^k) E_{G-band}$, where the $k$-independent term $\delta_0$ accounts for the combined effect of impurity scattering and Fermi velocity distribution. Reasonable matches with experimental data for the $B$-dependent component of the $G$ band are obtained using $\delta_0 = 6.3$ meV and $\Delta B = 1.7$ T for sample S1 (solid line in Fig.~\ref{SLGB}) and $\delta_0 = 6.3$ meV and $\Delta B = 1.0$~T for sample S2 (solid line in Fig. \ref{fig:SLG2_sim}). This result is consistent with the hypothesis that the LL broadening that damps the resonances at low fields in single-layer samples indeed arises from inhomogeneous strain fields that are not present in the MEG sample. Indeed, it is known that the pseudomagnetic fields associated with such strain fields could reach values up to tens of teslas in extreme cases;\cite{Yeh20111649} therefore the obtained $\Delta B = 1.0$ and $1.7$~T for our single layer samples on SiO$_2$ are reasonable values. Also, the results of magnetophonon resonance on single-layer graphene encapsuled on hexagonal boron nitride by Neumann {\it et al.} could be fit only by using increasing $\delta_k$ parameters for increasing transition indexes $k$.\cite{neumann2015magneto} Applying our model to those parameters one could conclude that their encapsulated sample yields $\Delta B \approx 0.4$~T, significantly smaller than for our samples deposited on SiO$_2$, as expected.

Finally, we should mention that the MPR resonances may also be influenced by sample-dependent electron-electron Coulomb interactions in a nontrivial way.\cite{Faugeras2015} Indeed, these interactions might be responsible for offsets in the MPR fields with respect to those given by the one-electron Dirac equation. However, these interactions, if homogeneous, are not expected to account for the washing out of the low-field magnetophonon resonances in the single-layer graphene samples studied here. On the other hand, inhomogeneities on electron-electron interactions are possible causes of Landau level broadening, which would likely be interconnected with the inhomogeneous strain fields. This would potentialize even further the influence of the latter on damping the Landau levels at low fields.

\section{Conclusions}

In summary, a comparative analysis of the magnetophonon resonances in single- and multilayer graphene samples indicated an additional extrinsic LL broadening mechanism for single-layer (and possibly few-layer) graphene associated with inhomogeneous strain. This mechanism becomes more important at lower magnetic fields.

\begin{acknowledgments}
We thank M. A. Pimenta, E. Nery, F. Plentz, L. Malard and L. C. Campos for helpful discussions. This work was supported by CAPES, FAPESP, FAPEMIG, CNPq, Nanocarbon INCT, and the Nanofabrication Network, Brazil.
\end{acknowledgments}

\end{document}